\newcommand{\p}[1]{(\ref{#1})}

\newcommand{\nn}{\nonumber}

\def\hybrid{\topmargin 0pt      \oddsidemargin 0pt
        \headheight 0pt \headsep 0pt
        \textwidth 160true mm       
        \textheight 231true mm         
        \marginparwidth 0.0in
        \parskip 0pt plus 1pt   \jot = 1.5ex}
\catcode`\@=11
\def\marginnote#1{}

\newcount\hour
\newcount\minute
\newtoks\amorpm
\hour=\time\divide\hour by60
\minute=\time{\multiply\hour by60 \global\advance\minute by-\hour}
\edef\standardtime{{\ifnum\hour<12 \global\amorpm={am}%
        \else\global\amorpm={pm}\advance\hour by-12 \fi
        \ifnum\hour=0 \hour=12 \fi
        \number\hour:\ifnum\minute<10 0\fi\number\minute\the\amorpm}}
\edef\militarytime{\number\hour:\ifnum\minute<10 0\fi\number\minute}

\def\draftlabel#1{{\@bsphack\if@filesw {\let\thepage\relax
   \xdef\@gtempa{\write\@auxout{\string
      \newlabel{#1}{{\@currentlabel}{\thepage}}}}}\@gtempa
   \if@nobreak \ifvmode\nobreak\fi\fi\fi\@esphack}
        \gdef\@eqnlabel{#1}}
\def\@eqnlabel{}
\def\@vacuum{}
\def\draftmarginnote#1{\marginpar{\raggedright\scriptsize\tt#1}}

\def\draft{\oddsidemargin -.5truein
        \def\@oddfoot{\sl preliminary draft \hfil
        \rm\thepage\hfil\sl\today\quad\militarytime}
        \let\@evenfoot\@oddfoot \overfullrule 3pt
        \let\label=\draftlabel
        \let\marginnote=\draftmarginnote
   \def\@eqnnum{(\theequation)\rlap{\kern\marginparsep\tt\@eqnlabel}%
\global\let\@eqnlabel\@vacuum}  }

\relax
\hybrid

\documentclass[12pt]{article}
\def\ba{\begin{eqnarray}}\def\ea{\end{eqnarray}}\def\lb{\label}
\def\bb{{\bf b}}\def\cc{{\bf c}}
\def\be{\begin{equation}}\def\ee{\end{equation}}\def\theequation{\thesection.\arabic{equation}}
\begin{document}
\thispagestyle{empty}
\vspace{2cm}
\vspace{3cm}
\begin{center}
{\Huge \bf BRST charges for finite nonlinear algebras}
\end{center}
\vspace{1cm}

\begin{center}
{\large\bf A.P.~Isaev${}^{a}$, S.O.~Krivonos${}^{a}$,
O.V.~Ogievetsky${}^{b}$ }
\end{center}

\begin{center}
${}^{a}$ Bogoliubov Laboratory of Theoretical Physics, \\
Joint Institute for Nuclear Research, \\
Dubna, Moscow region 141980, Russia \\
 isaevap@theor.jinr.ru, krivonos@theor.jinr.ru
\\
\vspace{0.3cm}
${}^{b}$ Center of Theoretical Physics\footnote{Unit\'e Mixte de Recherche
(UMR 6207) du CNRS et des Universit\'es Aix--Marseille I,
Aix--Marseille II et du Sud Toulon -- Var; laboratoire affili\'e \`a la
FRUMAM (FR 2291)}, Luminy,
13288 Marseille, France \\
and P. N. Lebedev Physical Institute, Theoretical Department,
Leninsky pr. 53, 117924 Moscow, Russia \\
 oleg@cpt.univ-mrs.fr
\end{center}
\vspace{1cm}

\begin{abstract}\noindent
Some ingredients of the BRST construction for quantum Lie algebras are applied to a wider class of 
quadratic algebras of constraints. We build the BRST charge for a quantum Lie algebra with three 
generators and ghost-anti-ghosts commuting with constraints. We consider a one-parametric family 
of quadratic algebras with three generators and show that the BRST charge acquires the 
conventional form after a redefinition of ghosts. The modified ghosts form a quadratic algebra. The family possesses a non-linear involution, which implies the existence of two independent BRST 
charges for each algebra in the family. These BRST charges anticommute and form a double BRST 
complex. 
\end{abstract}

\newpage\setcounter{page}{1}\setcounter{footnote}0

\setcounter{equation}0
\section{Introduction}
The construction of BRST charges $Q$ for linear (Lie) algebras of constraints is well known. In the case of nonlinear 
algebras, despite the existence of quite general results concerning the structure of the BRST charges (see, e.g., 
\cite{GTHT}, \cite{111}, \cite{IsOg4} and references therein), the general construction is far from being fully understood. 
The main reason is the appearance of non-standard terms in $Q$. Another issue is a possible existence of non-linear 
invertible transformation which preserves a certain form of relations (say, leaves the relations quadratic). The BRST charge 
might have a simple form in one basis while in other bases it becomes  cumbersome.
 
Among the quadratically nonlinear algebras there is a special class of so called quantum Lie 
algebras (QLA) (see \cite{IsOg1} -- \cite{IsOgGo} and references therein). The additional QLA
restrictions help to construct explicitly the BRST charges \cite{IsOg4,IsOgGo,IsOg3}. The main 
ingredient of the construction in \cite{IsOg4,IsOgGo,IsOg3} is the modified ghost-anti-ghost 
algebra which is also quadratically nonlinear. Moreover, in general, the ghost-anti-ghosts do not 
commute with the generators of the algebra. Unfortunately, the class of QLA's is not wide enough 
to include many interesting algebras. Therefore it seems desirable to extend at least some 
elements of the construction of the BRST charge for QLA to broader classes of quadratic algebras. 
Here we report on some preliminary results in this direction. In Sections 2 and 3 we relax 
one of the restrictions to make the algebra of constraints commute with the ghost-anti-ghosts. The 
BRST charges $Q$ can be built explicitly in this case. In Section 4 we discuss an example of 
QLA with three generators and present its BRST charge. In the next Section we 
construct BRST charges for a one-parametric family of quadratic algebras. Two nontrivial features arise. First, the BRST charge $Q$ takes a conventional form after a redefinition of the canonical 
ghost-anti-ghost system. The algebra of modified ghosts is quadratic as for QLA's. 
Second, the family admits a non-linear involution; it follows that any algebra of the family has 
two different bases with quadratic defining relations (two "quadratic faces") and therefore two
different BRST charges. It turns out that these BRST charges anticommute and form thus a double 
BRST complex.

\setcounter{equation}0
\section{Quantum space formalism}\setcounter{equation}0
Let $V_{N+1}$ be an $(N+1)$-dimensional vector space. Let $R\in {\rm End}(V_{N+1}\otimes V_{N+1})$
be a Yang-Baxter R-matrix, that is, a solution of the Yang-Baxter equation
\be\lb{rrr3}R_{\underline{2} \underline{3}}\, R_{\underline{1}\underline{2}}\, 
R_{\underline{2}\underline{3}}=R_{\underline{1}\underline{2}}\, R_{\underline{2}\underline{3}}\, 
R_{\underline{1}\underline{2}}\; \in {\rm End}(V_{N+1} \otimes V_{N+1}  \otimes V_{N+1}) \; \ee
(here $\underline{1},\underline{2}$ or $\underline{2},\underline{3}$ denote copies of the vector 
spaces $V_{N+1}$ on which the $R$-matrix acts nontrivially) or, in components $R^{AB}_{CD}$
($A,B,C,D=0,1,\dots,N$),
\be\label{rrr}R^{C_2 C_3}_{A_2A_3}\, R^{B_1D_2}_{A_1C_2}\, R^{B_2B_3}_{D_2C_3}=R^{C_1C_2}_{A_1A_2} 
\, R^{D_2B_3}_{C_2A_3} \, R^{B_1B_2}_{C_1 D_2} \; .\ee

Consider an algebra with generators $\chi_A=\{\chi_0,\chi_i\}$ $(i=1,\dots,N)$ and quadratic 
relations
\be\label{rtt1}R^{CD}_{AB}\,\chi_C\,\chi_D=\chi_A\,\chi_B\qquad {\mathrm{or}}\qquad
(1-R_{\underline{12}})\,\chi_{\underline{1}\rangle}\,\chi_{\underline{2}\rangle}=0\; .\ee
This algebra is usually called "quantum space" algebra. 

We extend the algebra (\ref{rtt1}) by ghosts $c^A$ with the following commutation relations
with $\chi_A$
\be\lb{qsp1}\chi_A c^D=c^B F^{CD}_{BA}\,\chi_C\; .\ee
Here $F$ is another Yang-Baxter matrix,
\be\lb{fff3}F_{\underline{2}\underline{3}}\, F_{\underline{1}\underline{2}}\, 
F_{\underline{2}\underline{3}}=F_{\underline{1}\underline{2}}\, F_{\underline{2}\underline{3}}\, 
F_{\underline{1}\underline{2}}\; ,\ee
which is compatible with the matrix $R$ in the sense that
\be\lb{frrf}R_{\underline{2}\underline{3}}\, F_{\underline{1}\underline{2}}\, 
F_{\underline{2}\underline{3}}=F_{\underline{1}\underline{2}}\, F_{\underline{2}\underline{3}}\, 
R_{\underline{1}\underline{2}}\quad ,\quad F_{\underline{2}\underline{3}}\, 
F_{\underline{1}\underline{2}}\, R_{\underline{2}\underline{3}}=R_{\underline{1}\underline{2}}\, 
F_{\underline{2}\underline{3}}\, F_{\underline{1}\underline{2}}\; .\ee
The matrix $F$ is called "twisting" matrix for the Yang-Baxter matrix $R$ (eqs.(\ref{rrr3}), 
(\ref{fff3}) and (\ref{frrf}) imply that the twisted matrix $\widetilde{R}=FRF^{-1}$ satisfies the 
Yang-Baxter equation as well).

The multiplication of ghosts is "wedge" with respect to the matrix $\widetilde{R}$; for 
quad\-ra\-tic combinations it reads
\be\lb{qsp3}c^Jc^I:=c^D\otimes c^B(1^{IJ}_{BD}-\widetilde{R}^{IJ}_{BD})\quad ,\quad 
1^{IJ}_{BD}:=\delta^{I}_{B}\delta^{J}_{D}\ .\ee

The algebra (\ref{rtt1}), (\ref{qsp1}) and (\ref{qsp3}) is graded by the ghost number:
${\rm gh}(\chi_A)=0$, ${\rm gh}(c^A)=+1$.

The element
\be\lb{brst2}Q:=c^A\,\chi_A\equiv c^i\,\chi_i+c^0\,\chi_0\ee
can be interpreted as a BRST operator for the quantum space algebra (\ref{rtt1}). Indeed, using 
(\ref{rtt1}), (\ref{qsp1}) and (\ref{qsp3}) one checks that $Q^2=0$,
$$\begin{array}{c}Q^2=c^{\langle\underline{2}}\,\chi_{\underline{2}\rangle}\, c^{\langle
\underline{2}}\,\chi_{\underline{2}\rangle}=c^{\langle\underline{2}}c^{\langle\underline{1}}\, 
F_{\underline{12}}\chi_{\underline{1}\rangle}\,\chi_{\underline{2}\rangle}\\[1em]
=c^{\langle\underline{2}}\otimes c^{\langle\underline{1}}\, (1-\widetilde{R}_{\underline{12}})\, 
F_{\underline{12}}\,\chi_{\underline{1}\rangle}\,\chi_{\underline{2}\rangle}=c^{\langle 
\underline{2}}\otimes c^{\langle\underline{1}}\, F_{\underline{12}}\, (1-R_{\underline{12}})
\chi_{\underline{1}\rangle}\,\chi_{\underline{2}\rangle}=0 \; .\end{array}$$

In the next Section we will consider the special choice of Yang-Baxter matrices $R$ and $F$ for 
which the generator $\chi_0$ is a central element for the algebra (\ref{rtt1}) and (\ref{qsp1}). 
In this case one can fix $\chi_0=1$ and then represent the ghost variable $c^0$ as a series 
\be\lb{cc0}c^0=\sum_{k=1}\sum_{i_\alpha,j_\beta =1}^N\, c^{i_{k+1}}\otimes\cdots\otimes c^{i_{1}} 
X_{i_{1}\dots i_{k+1}}^{j_{1}\dots j_{k}} b_{j_1}\cdots b_{j_k}\;\;\;\; \left({\rm gh}(c^0)=+1 
\right) \; ,\ee
where $X_{i_{1} \dots i_{k+1}}^{j_{1} \dots j_{k}}$ are constants and $b_A=\{ b_0,b_i\}$ 
anti-ghost generators with the ghost number ${\rm gh}(b_A)=-1$. The anti-ghosts $b_A$ satisfy
\be\lb{qsp2}b_A\chi_B=F^{CD}_{AB}\chi_C\, b_D \; ,\ee
\be\lb{qsp4}b_Ab_B=(1^{IJ}_{AB}-\widetilde{R}^{IJ}_{AB})b_I\otimes b_J\; ,\ee
\be\lb{qsp5}b_A\, c^B=-c^D\, (\widetilde{R}^{-1})^{CB}_{DA}\, b_C+D^B_A\; ,\ee
where $D^B_A$ is a constant matrix such that $D^B_0=0$, $D^i_j=\delta^i_j$. The compatibility of $c^0$ (\ref{cc0}) with 
(\ref{qsp1}), (\ref{qsp3}) and (\ref{qsp5}) yields the unique solution for tensors
$X_{i_{1}\dots i_{k+1}}^{j_{1}\dots j_{k}}$ in terms of the matrix components $F^{CD}_{AB}$ and
$R^{CD}_{AB}$. In papers \cite{IsOg4}, \cite{IsOgGo} we analyzed the case $F=R$
with a particular $R$-matrix (see eq.(\ref{rmat1}) below) and found in this case the unique solution
\be\label{solu7}\!\!\!\!\!\! X_{i_{1}\dots i_{r+1}}^{j_{1}\dots j_{r}}\!\! =\!\! (-1)^{r+1}\!
\left(\! (1-R^2_{\underline{r}})\! (1+R_{\underline{r-1}}R^2_{\underline{r}})\!\dots\!
(1+(-1)^{r}R_{\underline{1}}\!\dots\! R_{\underline{r-1}}R_{\underline{r}}^2)\!
\right)_{i_{1}\dots i_r,i_{r+1}}^{j_{1} \dots j_{r},\; 0}\ee
where $R_{\underline{k}}:=R_{\underline{k,k+1}}$ and $i_k,j_m=1,2,\dots,N$. In next Sections we 
will investigate examples of quadratic algebras (\ref{rtt1}), (\ref{qsp1}) and (\ref{qsp3}) with 
$F\neq R$. 

\setcounter{equation}0
\section{BRST operator for finitely generated quadratic algebras}

Consider a $(N+1)^2\times (N+1)^2$ Yang-Baxter matrix with the following restrictions on the 
components $R^{CD}_{AB}$ \cite{Ber}:
\be\label{rmat1}R^{ij}_{kl}=\sigma^{ij}_{kl}\; ,\;\;\; R^{0j}_{kl}=C^{j}_{kl}\; ,\;\;\; R^{0A}_{B0}=R^{A0}_{0B}=\delta^A_B\;\ee
(other components of $R$ vanish). Small letters $i,j,k,\dots =1,\dots ,N$ denote indices of the
$N$-dimensional subspace $V_N\subset V_{N+1}$.

For the $R$-matrix of the special form (\ref{rmat1}), the relations (\ref{rtt1}) are equivalent to
$$[\chi_0,\,\chi_i]=0\ \ {\rm and}\ \ (1-\sigma_{12})\,\chi_{1\rangle}\,\chi_{2\rangle }= 
C_{12\rangle}^{\langle 1}\,\chi_0\,\chi_{1\rangle }\ .$$
The generator $\chi_0$ is central and one can rescale the remaining generators, 
$\chi_i\rightarrow\chi_0\, \chi_i$. The rescaled generators (still denoted by $\chi_i$, $i =1,2,\dots,N$) satisfy relations 
\be\lb{qlaK}\chi_{i_1}\,\chi_{i_2}-\sigma^{k_1 k_2}_{i_1 i_2}\,\chi_{k_1}\,\chi_{k_2}=
C^{k_1}_{i_1i_2}\,\chi_{k_1} \qquad {\mathrm{or}}\qquad\chi_{1\rangle}\,\chi_{2\rangle}-\sigma_{12}
\,\chi_{1\rangle}\,\chi_{2\rangle}=C^{\langle 1}_{12\rangle}\,\chi_{1 \rangle} \; .\ee

For the R-matrix (\ref{rmat1}) the Yang-Baxter equation (\ref{rrr3}) imposes certain conditions 
for the structure constants $\sigma^{ij}_{kl}$ and $C^k_{ij}$ which can be written in the concise 
matrix notation \cite{FRT}, \cite{IsOg2} as
\be\label{int1a}\sigma_{12}\,\sigma_{23}\,\sigma_{12}=\sigma_{23}\,\sigma_{12}\,\sigma_{23}\; ,\ee
\be\label{int2}\begin{array}{c}C^{\langle 1}_{12\rangle}\, C^{\langle 4}_{13\rangle}=
\sigma_{23}\, C^{\langle 1}_{12\rangle}\, C^{\langle 4}_{13\rangle}+C^{\langle 3}_{23\rangle}\, 
C^{\langle 4}_{13\rangle}\; ,\end{array}\ee
\be\label{int3}C^{\langle 1}_{12\rangle}\,\sigma_{13}=\sigma_{23}\,\sigma_{12}\, 
C^{\langle 3}_{23\rangle}\; ,\ee
\be\label{int3a}\begin{array}{c}(\sigma_{23}\, C^{\langle 1}_{12\rangle}+
C^{\langle 3}_{23\rangle })\,\sigma_{13}=\sigma_{12}\, (\sigma_{23}\, C^{\langle 1}_{12\rangle }+
C^{\langle 3}_{23\rangle }) \; .\end{array}\ee
The condition (\ref{int1a}) says that $\sigma$ is the braid (Yang-Baxter) matrix, condition 
(\ref{int2}) is a version of the Jacobi identity. The quadratic algebra (\ref{qlaK}) with 
conditions (\ref{int1a}) -- (\ref{int3a}) is called {\it quantum Lie algebra} (QLA). The usual Lie 
algebras form a subclass of the QLA corresponding to $\sigma^{ij}_{km} = \delta^i_m \delta^j_k$
(i.e., $\sigma$ is the permutation).

Below we consider the simplest, unitary, braid matrices $\sigma$, that is,
\be\lb{s2eq1}\sigma_{nm}^{pj}\sigma^{ki}_{pj}=\delta^k_n\delta^i_m\qquad {\mathrm{or}}\qquad
\sigma^2 = 1\; .\ee
Then (\ref{int3a}) follows from (\ref{int3}) and symmetries of (\ref{qlaK}) imply that
\be\lb{add2}(1+\sigma_{12})C^{\langle 1}_{12\rangle}=0 \; .\ee

The generators $c^i$, $b_i$ $(i=1,\dots,N)$ of the ghost-anti-ghost algebra satisfy quadratic 
relations
\be\lb{qlabc}b_{1\rangle}\, b_{2\rangle}=-\tilde{\sigma}_{12}\, b_{1\rangle}\, b_{2\rangle}\; , 
\;\;\; c^{\langle 2}\, c^{\langle 1}=-c^{\langle 2}\, c^{\langle 1}\tilde{\sigma}_{12}\; ,\ee
\be\lb{qlabcK}b_{2\rangle}\, c^{\langle 2}=-c^{\langle 1}\,\tilde{\sigma}^{-1}_{12}\, 
b_{1\rangle}+I_2\; ,\ee
where $\tilde{\sigma}_{12}=\phi_{12}\sigma_{12}\phi_{12}^{-1}$. These relations are obtained from 
eqs.(\ref{qsp3}), (\ref{qsp4}) and (\ref{qsp5}) for the special choice of the matrix $F$:
\be\label{fmat1}F^{ij}_{kl}=\phi^{ij}_{kl}\; ,\;\;\; F^{0A}_{B0}=F^{A0}_{0B}=\delta^A_B \; ,\ee 
(other components vanish).

A cross-product of the QLA (\ref{qlaK}) and the ghost algebra (\ref{qlabc}), (\ref{qlabcK})
is defined by the commutation relations (\ref{qsp1}) and (\ref{qsp2}),
\be\label{crossKK}b_{1\rangle}\,\chi_{2\rangle}=\phi_{12}\,\chi_{1\rangle}\, b_{2\rangle }\; , 
\;\;\;\chi_{2\rangle}\, c^{\langle 2}=c^{\langle 1}\,\phi_{12}\,\chi_{1\rangle}\; .\ee
We denote this cross-product algebra by $\Omega$. For consistency of the algebra $\Omega$ we 
require that the matrix $\phi$ satisfies relations
\be\lb{cons1}\begin{array}{c}\sigma_{12}\,\phi_{23}\,\phi_{12}=\phi_{23}\,\phi_{12}\,\sigma_{23} 
\; , \;\;\phi_{12}\,\phi_{23}\,\sigma_{12}=\sigma_{23}\,\phi_{12}\,\phi_{23}\; ,\\[1em]
\phi_{12}\,\phi_{23}\,\phi_{12}=\phi_{23}\,\phi_{12}\,\phi_{23}\; ,\;\;\end{array}\ee
\be\lb{cons3}\phi_{12}\phi_{23}C_{12\rangle}^{\langle 1}\delta_{3\rangle}^{\langle 2}=
C_{23\rangle}^{\langle 2}\,\phi_{12}\; ,\ee
which follow from relations (\ref{frrf}) and (\ref{fff3}) with the Yang-Baxter matrices $R$ and 
$F$ given by (\ref{rmat1}) and (\ref{fmat1}).

Now the construction (\ref{brst2}) of the BRST operator for the QLA (\ref{qlaK}) and the ghost 
algebra (\ref{qlabc}), (\ref{qlabcK}),  (\ref{crossKK}) gives the following result \cite{iskrog}:

\vspace{0.1cm}\noindent
{\bf Proposition.} {\it Let $c^0 = - \frac{1}{2} \, c^{j} \, c^{i} \; \phi_{ij}^{km} C^{r}_{km}  
\, b_{r}$. Then the element $Q\in\Omega$, 
\be\lb{brstK}Q=c^{j}\chi_{j}+c^0\in\Omega\ee
satisfies
\be\lb{zero}Q^2 = 0 \; .\ee}

For a general braid matrix $\sigma$, there are always two possibilities for the twisting matrix 
$\phi$. The first possibility is $\phi =\sigma$; it was investigated in 
\cite{IsOg1}--\cite{IsOg3}. The second one is $\phi^{kl}_{nm}=\delta^k_m\delta^l_n$ which leads to 
the tensor product of the algebra (\ref{qlaK}) and the ghost-anti-ghost algebra 
(\ref{qlabc}), (\ref{qlabcK}). In other words, with this choice, the ghosts commute with the 
generators of the QLA,
\be\label{crossK}b_i\,\chi_j=\chi_j\, b_i\; ,\;\;\; c^i\,\chi_j=\chi_j\, c^i\; .\ee
This possibility will be considered in the next Sections on examples of 3-dimen\-sional nonlinear 
algebras.

\setcounter{equation}0
\section{Example of a 3-dimensional QLA}
In this Section we present an explicit example of a finite-dimensional QLA 
(\ref{qlaK})--(\ref{s2eq1}) and construct the BRST charge for this algebra.

The algebra we start with has four generators $\{\chi_0,\chi_1,\chi_2,\chi_3\}$ which obey the 
following quadratic relations
\be\lb{qlaf1}\begin{array}{c}[\chi_1,\,\chi_2]=0\; ,\;\;\; [\chi_1,\,\chi_3]=\alpha\chi_1^2+ 
\chi_0\,\chi_2\; ,\;\;\; [\chi_2,\,\chi_3]=\alpha\,\chi_1\,\chi_2\; ,\\ [0.2cm]
[\chi_0,\,\chi_i]=0\ \ ,\ i=1,2,3 \; ,\end{array}\ee
where $\alpha\neq 0$ is a parameter. This parameter can be set to one, $\alpha=1$ 
by rescaling of the generators $\chi_A$. One can write (\ref{qlaf1}) in the form (\ref{rtt1})
with the $R$-matrix
\be\lb{qlaf3}\begin{array}{c}R^{AB}_{CD}=\delta^A_D\,\delta^B_C+\left(\delta^A_0\,\delta^B_2+  
\alpha\delta^A_1\,\delta^B_1\right)\left(\delta^1_C\,\delta^3_D-\delta^3_C\,\delta^1_D\right)+
\\[0.2cm]
+\alpha\left(\delta^A_1\,\delta^B_2\,\delta^2_C\,\delta^3_D-\delta^A_2\,\delta^B_1\,\delta^3_C\, 
\delta^2_D\right) \; .\end{array}\ee
The matrix (\ref{qlaf3}) satisfies the Yang-Baxter equation; it is of the form
(\ref{rmat1}) with
\be\label{rmat11}\begin{array}{l}\sigma^{ij}_{kl}=\delta^i_l\,\delta^j_k+\alpha\,\delta^i_1\,
\delta^j_1\left(\!\delta^1_k\,\delta^3_l-\delta^3_k\,\delta^1_l\!\right) +\alpha\left(
\!\delta^i_1\,\delta^j_2\,\delta^2_k\,\delta^3_l-\delta^i_2\,\delta^j_1\,\delta^3_k\,\delta^2_l\!
\right) ,\\[.5em]
C^{j}_{kl}=\delta^i_0\,\delta^j_2\left(\delta^1_k\,\delta^3_l-\delta^3_k\,\delta^1_l\right)\ \ 
,\ i,j,k,l=1,2,3 \; . \end{array}\ee
The matrix $\sigma$ has the form $\sigma_{12}=P_{12}+u_{12}$, where $u_{12}=-u_{21}$ and 
$u_{12}^2=0$, so $\sigma^2=1$ ($\sigma$ belongs to the family F in the classification of GL(3)
R-matrices in \cite{EO}). Thus, for $\chi_0=C=$const, the algebra (\ref{qlaf1}) is an example of 
the QLA (\ref{qlaK})--(\ref{s2eq1}).

According to the choice of the structure constants, the non-canonical ghost-anti-ghost algebra (\ref{qlabc}), (\ref{qlabcK}) and (\ref{crossK}) reads:
\be\label{exam2}\begin{array}{c}
(c^1)^2=\alpha c^3\, c^1\; ,\;\;\; (c^2)^2=(c^3)^2=0\; ,\;\;\; \{ c^1,\, c^3\}=\{c^2,\,c^3\}=0\; , 
\;\;\; \{c^1,\, c^2\} =\alpha c^3\,c^2\; ,\\[0.2cm] 
(b_1)^2=(b_2)^2=(b_3)^2=0\; ,\;\;\; \{ b_1,\, b_2\} =\{ b_1,\, b_3\} =0\; ,\;\;\;\{ b_2,\, b_3\}= 
\alpha b_1b_2\; ,\\[0.2cm] 
\{ b_1,\, c^1\} =-\alpha c^3\, b_1+1,\;\{b_2,\, c^2\} =\{b_3,\, c^3\} =1,\;\{ b_3,\, c^2\} = 
\alpha c^2\, b_1\, ,\\[0.2cm] 
\{ b_2,\, c^1 \} =-\alpha c^3\, b_2\, ,\;\;\{ b_3,\, c^1\} =\alpha c^1\, b_1\, ,\;\; 
\{ b_1,\, c^2\} =\{b_1,\, c^3\} =\{b_2,\, c^3\} =0\, ,\\[0.2cm] 
\, [\chi_i,\, c^j]=0=[\chi_i,\, b_j]\; .\end{array} \ee  
where $\{.,.\}$ stands for the anti-commutator. Then the BRST operator (\ref{brstK}) for the 
ghost-anti-ghost algebra (\ref{exam2}) has the standard form
\be\label{exam3}Q=\sum_{i=1}^3c^i\chi_i-c^1\, c^3\, C\, b_2\; , \ee
and one can recheck directly that $Q^2=0$.

We note that under the following nonlinear invertible transformation of the generators,
$$\chi_2\mapsto\chi_2+\gamma\chi_1^2\; ,$$
where $\alpha = 2 \gamma \chi_0$, the relations (\ref{qlaf1}) have a different, but still 
quadratic, form
\be\lb{qlaf11}[\chi_1,\,\chi_2]=0\; ,\;\;\; [\chi_1,\,\chi_3]=\frac{\alpha}{2}\,\chi_1^2+\chi_0\, 
\chi_2\; ,\;\;\; [\chi_2,\,\chi_3]=2\alpha\,\chi_1\,\chi_2\; .\ee
These relations cannot be presented in the form (\ref{rtt1}) with an $R$-matrix (\ref{rmat1})
and any GL(3) matrix $\sigma$.

{}For the ghost algebra (\ref{exam2}) the Fock space $F$ is constructed in the standard way. Let 
$V$ be a left module over the algebra (\ref{qlaf1}). For any vector $|\psi\rangle\in V$ we require
\be\label{exam33}b_i |\psi\rangle =0\; ,\ee
i.e., the anti-ghosts $\{ b_i \}$ are annihilation operators for all vectors in $V$. Then the Fock 
space $F$ is generated from $V$ by the ghost operators $\{ c^i \}$  (creation operators) and in 
view of (\ref{exam2}) any vector  $| \Phi \rangle \in F$ has the form
\be\label{exam4}|\Phi\rangle =|\psi_0\rangle +\sum_{i=1}^3c^i|\psi_i\rangle +\sum_{i<j}c^i c^j| 
\psi_{ij}\rangle +c^1c^2c^3|\psi_{123}\rangle\; ,\ee
where $|\psi_{\dots}\rangle\in V$. The "physical subspace" in $F$ is extracted by the condition
\be\label{exam5}Q|\Phi\rangle =0\; ,\ee
which gives
$$\chi_i|\psi_0\rangle =0\ \ ,\ i=1,2,3\; ,\;\;\;\dots\; .$$
Since the vector $| \psi_0 \rangle$ is annihilated by the first class constraints $\chi_i$, this 
vector belongs to the physical subspace in $V$.

In the next Section we will show that the quadratic ghost algebra (\ref{exam2}) can be realized 
in terms of the canonical ghosts and anti-ghosts $\{ {\bf c}^i, {\bf b}_j \}$ (cf. the standard
deformation of the algebra of the bosonic creation and annihilation operators, \cite{Og}).

\setcounter{equation}0
\section{BRST operator for a 3-dimensional nonlinear algebra}

We construct the BRST operator for the algebra, which generalizes the QLAs (\ref{qlaf1}) and (\ref{qlaf11}):
\be\lb{eq11}[J,\, W]=a_1T+a_2J^2\; ,\;\;\; [J,\, T]=0 \; ,\;\;\; [T,\, W]=a_3J\, T\; ,\ee
with $a_1,a_2,a_3\neq 0$. By rescaling of the generators, two of three coefficients $\{a_1,a_2\}$ 
or $\{a_1,a_3\}$ may be arbitrarily fixed. In what follows we prefer to leave all these 
coefficients free and fix them at the end of calculations, if needed.
 
The values $a_3/a_2 =1$ (respectively, $a_3/a_2 =4$) correspond to the algebra (\ref{qlaf1}) 
(respectively, (\ref{qlaf11})), where we should identify
$$\chi_1=J\; ,\;\;\;\chi_2=T\; ,\;\;\;\chi_3=W\; .$$
For $a_3/a_2 =-16$ or $a_3/a_2 =-1/4$ this algebra is a finite dimensional "cut" of the bosonic 
part of the $N=2$ super $W_3$ algebra  \cite{ikm} $\{ J=J_{-1},T={\tilde L}_{1},W=W_{2}\}$.

The algebra \p{eq11} is quadratic and we may construct the BRST charge using quadratic ghosts 
along the lines discussed in the beginning of this paper (see \cite{IsOg1} -- \cite{iskrog} also). 
Nevertheless, to make steps more transparent we first construct the BRST charge with the 
canonical ghost-anti-ghost generators and then define the nonlinear ghosts systems in which the 
BRST charge drastically simplifies. So we introduce the fermionic ghost-anti-ghost generators 
$\{\bb_J,\cc^J,\bb_T,\cc^T,\bb_W,\cc^W\}$ with the standard relations
\be\label{ghosts}\left\{\bb_J,\cc^J\right\} =1,\quad\left\{\bb_T,\cc^T\right\} =1,\quad\left\{ 
\bb_W,\cc^W\right\} =1\ee
(other anti-commutators are zero).

By virtue of a rather simple structure of the algebra \p{eq11} the BRST charge can be easily found to be
\be\label{q11}Q=\cc^J\, J+\cc^{T}\, T+\cc^{W}\, W-a_1\cc^J\cc^{W}\bb_T-a_3T\,\cc^T\cc^{W}\bb_J+ 
a_2\, J\cc^W\cc^J\bb_J\; ,\ee
where we assumed the "initial condition" 
\be\lb{inicon}Q=\cc^J\, J+\cc^{T}\, T+\cc^{W}\, W+{\rm higher}\; {\rm order}\; {\rm terms}\ee
and used the ordering with the annihilation operators $\{ \bb_i \}$ on the right. If we relax the 
"initial condition" then the BRST charge is not unique. E.g., the operator
$Q'=Q+\mu J\,\cc^W$
($\mu$ is a constant) satisfies $(Q')^2=0$ as well. 

The last two terms in the BRST charge \p{q11} are unconventional. Let us now rewrite the BRST 
charge as follows
\be\label{q1a}\begin{array}{l}Q=\left(\cc^J+a_2\,\cc^W\cc^J\bb_J\right) J+\cc^{W}\, W-a_1\left( 
\cc^J+a_2\,\cc^W\cc^J\bb_J\right)\,\cc^{W}b_T\\[1em]  
\hspace{.8cm}+\left(\cc^{T}-a_3\cc^T\cc^{W}\bb_J\right) T\;.\end{array}\ee
{}It is now clear the BRST charge \p{q1a} acquires the conventional form (of the type 
(\ref{brstK})) after introducing "new" ghosts $\{c^J, c^T, c^W \}$: 
\be\label{gh1}c^J=\cc^J+a_2\,\cc^W\cc^J\bb_J,\quad c^T=\cc^{T}-a_3\cc^T\cc^{W}\bb_J,\quad c^W= 
\cc^{W} \; .\ee
In terms of new ghosts the BRST charge \p{q1a} reads
\be\label{q1b}Q=c^J\,J+c^{T}\,T+c^{W}\,W-a_1c^Jc^{W}\bb_T\; ,\ee
in agreement with the ideas discussed above and in \cite{IsOg1} -- \cite{iskrog}. It is 
straightforward to write the relations for the new ghost-anti-ghost generators (\ref{gh1}); they 
form a quadratic algebra
\begin{eqnarray}\lb{ccbb}&&\left\{ c^J,c^J\right\} =-2a_2c^J\, c^W,\quad\left\{ c^J,c^T\right\} 
=-a_3c^T\,c^W,\quad\left\{c^J,\bb_J\right\} =1-a_2\, c^W\bb_J,\nn\\
&&\left\{c^J,\bb_W\right\}=a_2\,c^J\bb_J,\quad\left\{c^T,\bb_T\right\}=1-a_3\,c^W\bb_J,\quad
\left\{c^T,\bb_W\right\}=a_3\,c^T\bb_J,\nn\\
&&\left\{c^W,\bb_W\right\}=1\; ,\end{eqnarray}
other anti-commutators are zero. To relate this ghost-anti-ghost algebra and the BRST charge 
(\ref{q1b}) with the algebra (\ref{exam2}) and the BRST charge (\ref{exam3}) we need also to 
redefine the anti-ghost variables
$$b_J=\bb_J\; ,\;\;\; b_T=\bb_T+a_3\cc^W\bb_J\bb_T\; ,\;\;\; b_W=\bb_W \; ,$$
and fix $a_2=a_3=\alpha$, $a_1=C$.

Thus, we see that the price we have to pay for the conventional form of the BRST charge is the 
quadratically nonlinear ghost-anti-ghost algebra, as it has been claimed in 
\cite{IsOg1}-\cite{IsOgGo}, \cite{iskrog}.
 
\subsection{Double BRST complex}
An interesting peculiarity of the family (\ref{eq11}) of non-linear algebras is an existence
of a non-linear redefinitions of the generators. Redefine the generator $T \mapsto {\cal T}$,
\be\label{TN}{\cal T}=T+\beta J\,J \ee
($\beta$ is a constant). In terms of generators $\{ J,{\cal T},W \}$ the algebra (\ref{eq11}) 
becomes cubic for general $\beta$. However, it is amusing that for 
\be\label{alpha}\beta=\frac{2a_2-a_3}{2a_1}\ee
the commutators of the generators $\{ J,{\cal T},W\}$ are again quadratic,
\be\lb{eq111}[J,\,W]=\tilde{a}_1{\cal T}+\tilde{a}_2J^2\; ,\;\;\; [J,\, {\cal T}]=0\; ,
\;\;\; [{\cal T},\, W]=\tilde{a}_3 J\, {\cal T}\; ,\ee
where
\be\label{tildea}{\tilde a}_1=a_1,\quad {\tilde a}_2=\frac{a_3}{2},\quad {\tilde a}_3=2a_2.\ee
By rescalings, one can set $a_1$ to 1 and leave $t=2a_2/a_3$ as the essential
parameter of the family. The transformation (\ref{tildea}) is the involution $\tilde{t}=1/t$.

Therefore, our (in general cubic) algebra has two "quadratic faces". Now we immediately conclude 
that for the second "face" another BRST charge ${\widetilde Q}$ exists, 
\be
\label{2BRST}{\widetilde Q}=\cc^J\, J+\cc^{T}\, {\cal T}+\cc^{W}\, W-{\tilde a}_1\cc^J\cc^{W} 
\bb_T-{\tilde a}_3{\cal T}\,\cc^T\cc^{W}\bb_J+{\tilde a}_2\, J\cc^W\cc^J\bb_J\; 
\ee
(it is constructed in the same way as \p{q11}). Moreover, one checks that
\be{\widetilde Q}{}^2=0,\quad\mbox{and}\quad\left\{ Q,{\widetilde Q}\right\}=0.\ee
Thus, for our algebra \p{eq11} we have two nontrivial BRST operators $Q ,{\widetilde Q}$ forming a 
double complex. Both operators are linear in the generators of the algebra and satisfy the initial 
condition (\ref{inicon}) but in different bases: $Q$ in the basis $\{ J,T,W\}$ and $\widetilde Q$ 
in the new basis $\{ J,{\cal T},W\}$. In the basis $\{ J,{\cal T},W\}$, the BRST operator $Q$ 
does not satisfy initial condition (\ref{inicon}), it contains nonlinear in $J$ terms. The same 
is true for $\widetilde Q$ in the basis $\{J,T,W\}$. 
{}For an algebra, having several quadratic faces, related by nonlinear transformations, one can 
impose standard initial condition in any of them and build -- in general nonequivalent -- BRST 
charges (cf. the Lie algebra $[x,y]=y$ and transformations $x\mapsto x+f(y)$, $f$ is a 
polynomial). 

  
\section{Conclusion}
We extended some elements of the construction of BRST charge for quantum Lie algebras to more 
general quadratic algebras. We explicitly found the BRST charges in the examples when the 
constraints commute with the ghost-anti-ghosts. We discussed an example of a QLA with three 
generators and presented the BRST charge for this algebra. As another interesting example we 
considered, as an analogue of a QLA, a one-parametric family of quadratic algebras with three 
generators. On this simple example we have shown that one can redefine the ghost-anti-ghost system 
in such a way that the BRST operator takes the conventional form $Q=c^i\chi_i+$"ghost terms" 
(\ref{brstK}). The modified ghosts form a quadratically nonlinear algebra as for QLA's. 
In addition, the members of this family admit two different presentations with quadratic defining 
relations. In agreement with general considerations in each presentation there is a 
conventional BRST charge. Being written in one basis they give rise to two inequivalent BRST 
charges $Q$, $\widetilde{Q}$ which anticommute and form a double BRST complex. We think that any 
algebra possessing several quadratic faces should have inequivalent BRST charges.
 
As immediate applications of our results one may try to construct the modified ghost-anti-ghosts system for some known 
nonlinear (super)algebras to simplify their BRST charges. Being extremely interesting 
(for us), this task seems to be less important than an analysis of situations with several BRST 
charges. 

\section*{Acknowledgements}
We are grateful to I.~Buchbinder and P.~Lavrov for valuable discussions.

\vskip .1cm
The work of A.P.I. was partially supported by the RF President Grant N.Sh.-195.2008.2 and by the 
grant RFBR-08-01-00392-a. The work of S.O.K. was partially supported by INTAS under contract 
05-7928 and by grants RFBR-06-02-16684, 06-01-00627-a, DFG~436 Rus~113/669/03. The work of O.V.O. 
was supported by the ANR project GIMP No.ANR-05-BLAN-0029-01.

\end{document}